\documentclass[]{mn2e}
\usepackage{graphicx}
\usepackage{epsfig}
\usepackage{amsmath}
\newcommand       \Angstrom     {\,{\rm \AA}}

\newcommand       \cm           {\,{\rm cm}}

\newcommand       \eV           {\,{\rm eV}}
\newcommand       \g            {\,{\rm g}}

\newcommand     \gtsim  {\lower.5ex\hbox{$\buildrel > \over \sim$}}
\newcommand     \ltsim  {\lower.5ex\hbox{$\buildrel < \over \sim$}}
\newcommand     \simgt  {\lower.5ex\hbox{$\buildrel > \over \sim$}}
\newcommand     \simlt  {\lower.5ex\hbox{$\buildrel < \over \sim$}}

\newcommand       \mum          {\,{\rm \mu m}}
\newcommand       \ppm          {\,{\rm ppm}}

\newcommand       \simali       {\sim\,}

\newcommand	  \NC           {N_{\rm C}}
\newcommand	  \NH           {N_{\rm H}}

\newcommand	  \epsre        {\varepsilon^{\prime}}
\newcommand	  \epsim        {\varepsilon^{\prime\prime}}
\newcommand	  \CTOHgraphene {\left[{\rm C/H}\right]_{\rm graph}}
\title[How much graphene in space?]{How much graphene in space?}
\author[Li, Li, \& Jiang]
       {Qi~Li$^{1,2}$,
        Aigen Li$^{2}$\thanks{lia@missouri.edu}, and
        B.W.~Jiang$^{1}$\thanks{bjiang@bnu.edu.cn}\\
        $^1$Department of Astronomy,
                Beijing Normal University,
                Beijing 100875, China\\
        $^2$Department of Physics and Astronomy,
             University of Missouri,
             Columbia, MO 65211, USA
             }
\begin{document}
\date{Accepted 2019 September 24; received 2019 September 10; in original form 2019 July 12.}

\pagerange{\pageref{firstpage}--\pageref{lastpage}} \pubyear{2019}

\maketitle

\label{firstpage}
\begin{abstract}
The possible presence of graphene 
in the interstellar medium (ISM) is examined by
comparing the interstellar extinction curve
with the ultraviolet absorption of graphene 
calculated from its dielectric functions experimentally
obtained with the electron energy loss spectroscopy 
(EELS) method. Based on the absence in the interstellar
extinction curve of the $\simali$2755$\Angstrom$
$\pi$--$\pi^{\ast}$ electronic interband transition 
of graphene, we place an upper limit of 
$\simali$20$\ppm$ of C/H 
on the interstellar graphene abundance,
exceeding the previous estimate
by a factor of $\simali$3 which made use of
the dielectric functions measured with 
the spectroscopic ellipsometry (SE) method.
Compared with the SE method 
which measures graphene in air 
(and hence its surface is contaminated)
in a limited energy range of $\simali$0.7--5$\eV$,
the EELS probes a much wider energy range of
$\simali$0--50$\eV$ and is free of contamination. 
The fact that the EELS dielectric functions are
substantially smaller than that of SE naturally 
explains why a higher upper limit on
the graphene abundance is derived with EELS.
Inspired by the possible detection of C$_{24}$, 
a planar graphene sheet, in several Galactic
and extragalactic planetary nebulae, 
we also examine the possible presence of
C$_{24}$ in the diffuse ISM 
by comparing the model IR emission
of C$_{24}$ with the observed IR emission
of the Galactic cirrus and the diffuse ISM
toward
$l = 44^{\rm o}20^\prime$ and 
$b=-0^{\rm o}20^\prime$.
An upper limit of $\simali$20$\ppm$
on C$_{24}$ is also derived from the absence
of the characteristic vibrational bands 
of C$_{24}$ at $\simali$6.6, 9.8 and 20$\mum$
in the observed IR emission.
\end{abstract}
\begin{keywords}
dust, extinction -- infrared: ISM --  ISM: lines and bands
           --- ISM: molecules
\end{keywords}

\section{Introduction}\label{sec:intro}
As the 4th most abundant element in the universe, 
carbon (C) is widespread in the interstellar medium (ISM)
in various allotropic forms (e.g., see Henning \& Salama 1998), 
including graphite (Groopman \& Nittler 2018),
nanodiamonds (Lewis et al.\ 1987, Allamandola et al.\ 1992,
Guillois et al.\ 1999, van Kerckhoven et al.\ 2002,
Shimonish et al.\ 2016),
fullerene (Cami et al.\ 2010, Sellgren et al.\ 2010, 
Garc{\'{\i}}a-Hern{\'a}ndez et al.\ 2010,
Zhang \& Kwok 2011, Bern\'e et al.\ 2013, 
Strelnikov et al.\ 2015, Bern\'e et al.\ 2017),
amorphous carbon and hydrogenated amorphous carbon 
(Wickramasinghe \& Allen 1980, 
Sandford et al.\ 1991,
Pendleton et al.\ 1994,
Greenberg et al.\ 1995,
Whittet et al.\ 1997,
Furton et al.\ 1999,
Pendleton \& Allamandola 2002,
Duley et al.\ 2005,
Dartois et al.\ 2007,
Chiar et al.\ 2013,
Mat{\'e} et al.\ 2016
G{\"u}nay et al.\ 2018),
and planar polycyclic aromatic hydrocarbons 
(PAHs; L\'eger \& Puget 1984, Allamandola et al.\ 1985).
Carbon buckyonions, composed of spherical concentric 
fullerene shells, are also postulated to be present in the ISM
(Chhowalla et al.\ 2003, Iglesias-Groth et al.\ 2003,  
Ruiz et al.\ 2005, Li et al.\ 2008, Qi et al.\ 2018).
The presence of various carbon allotropes in the ISM 
is revealed either through their fingerprint 
ultraviolet (UV) and/or infrared (IR) 
absorption or emission spectral features. 
Their presence in the ISM is also revealed through 
isotope analysis of primitive meteorites
as presolar grains of interstellar origin. 
Although the interstellar C abundance is not 
accurately known, it is generally believed that 
in the ISM $\simali$40--60\% of 
the interstellar C must have been tied up 
in these carbon allotropes 
(see Mishra \& Li 2015, 2017).

Graphene, a novel material experimentally
synthesized by A.K.~Geim and K.S.~Novoselov
(see Novoselov et al.\ 2004, 
but also see Henning et al.\ 2004), 
was first shown by 
Garc{\'{\i}}a-Hern{\'a}ndez et al.\ (2011b, 2012)
to be possibly present in the ISM based on 
the detection of the unusual IR emission features 
at $\simali$6.6, 9.8, and 20$\mum$ 
in several planetary nebulae (PNe), 
both in the Milky Way and in the Magellanic Clouds,
which are coincident with the strongest transitions 
of planar C$_{24}$, a piece of graphene.
Chen et al.\ (2017) argued that, in principle, 
graphene could be present in the ISM 
as it could be formed from
the photochemical processing of PAHs,
which are abundant in the ISM,
through a complete loss of 
their H atoms (e.g., see Bern\'e \& Tielens 2012).
Chuvilin et al.\ (2010) showed experimentally 
that C$_{60}$ could be formed from a graphene sheet.
Bern\'e \& Tielens (2012) further proposed that
such a formation route of converting PAHs 
into graphene and subsequently C$_{60}$
could occur in space, in the presence of UV irradiation.
More recently, Sarre (2019) suggested that graphene
oxide nanoparticles could be present in the ISM and
their photoluminescence could explain 
the widespread extended red emission (ERE),
a broad, featureless emission band 
between $\simali$5400$\Angstrom$
and $\simali$9500$\Angstrom$
(see Witt \& Vijh 2004, Witt 2014, 
Lai et al.\ 2017).

Based on the dieletric functions
measured in the wavelength range of 
$\simali$245--1700\,nm
(corresponding to $\simali$0.7--5\,eV)
by Nelson et al.\ (2010) 
for large-area polycrystalline 
chemical vapor deposited (CVD) graphene,
Chen et al.\ (2017) calculated the absorption 
of graphene from the UV to the near-IR.
By comparing the calculated absorption of graphene
with the observed interstellar extinction,
they found that as much as $\simali$5$\ppm$
of C/H could be tied up in graphene 
%without exhibiting the $\simali$2755$\Angstrom$ 
%or $\simali$4.5\,eV exciton-dominated absorption 
%peak of graphene which is absent 
%in the interstellar extinction.
without violating the observational 
constraints on the interstellar extinction
and IR emission.

%SE is a nondestructive optical technique 
%that allows for the measurement of a thin 
%film's thickness and/or dielectric function. 

However, the dielectric functions of graphene 
of Nelson et al.\ (2010) were derived from 
the spectroscopic ellipsometry (SE) method
in which the CVD graphene was exposed to air
and the majority of its surface was covered 
with adsorbates and contamination.
In order to discern the composition of 
the adsorbates and contamination, 
Nelson (2012) performed  an electron energy 
loss spectroscopy (EELS) measurement of 
the graphene sample over the core-loss 
energy region (i.e., hundreds of eV)
which detects inner-shell ionizations 
on the order of the atoms's binding energies. 
It was found that the EELS core-loss spectra
show the presence of oxygen, calcium and iron 
from exposure to the environment and the copper 
etchant/deionized water-based transfer process.\footnote{%
  In the chemical vapor deposition of graphene
  of Nelson et al.\ (2010), a hydrocarbon precursor 
  (methane) was delivered under low pressure 
  to a copper foil for deposition. 
  The CVD graphene was then transfered to
  a secondary SiO$_2$/Si glass substrate.
  The graphene transfer process consisted of
  applying a thermal release tape to the graphene 
  on copper foil, etching away the foil in an aqueous 
  solution of iron chloride, and applying 
  the graphene+tape to the substrate.  
  }

In contrast, the EELS method enables the isolation 
of the spectroscopy to a freely suspended area 
without contamination and therefore allows one
to measure the dielectric functions 
of perfectly clean graphene.
In EELS, inelastically scattered electrons are 
collected that have undergone Coulomb interaction 
with the atomic electrons of the sample 
and have been scattered to much smaller angles 
(on the oder of mrad).
The EELS over the low-loss region ($\simali$0--50\,eV) 
detects energy losses that are the result of plasmons,
interband transitions and intraband transitions. 
In comparison to the energy range probed by SE,
EELS extends this range by $\simali$6--10 times, 
depending on the bandwidth of the light source 
used in a comparative optical technique. 

Nelson et al.\ (2014) measured the EELS of 
monolayer CVD graphene in the low-loss 
energy region. They found that the EELS 
dielectric functions of graphene in the UV
and optical are substantially smaller than 
those measured by Nelson et al.\ (2010)
in terms of the SE technique.
Since the UV/optical absorption of graphene 
sensitively depends on its dielectric function,
a {\it smaller} dielectric function would lead to 
a {\it lower} UV/optical absorption level
(on a per unit mass basis).
By implication, it would also result in
a {\it smaller} amount of thermal emission
in the IR because of the energy balance 
between absorption and emission.
Therefore, the detection (or nondetection) of
any fingerprint spectral features
characteristic of graphene in the UV 
and IR would imply a {\it larger} amount 
of graphene in the interstellar space,
or place a {\it higher} upper limit 
in the case of nondetection. 
In this work, we re-visit the possible 
presence of graphene in the ISM
by applying the EELS-measured 
dielectric functions of graphene 
of Nelson et al.\ (2014)
and confronting the model extinction
(\S\ref{sec:extcurv}) and emission
(\S\ref{sec:irem}) with the observed
interstellar extinction curve and IR emission.
The major results are summarized 
in \S\ref{sec:summary}.

%Both quantum-chemical computations 
%and laboratory experiments have shown 
%that the %$\pi$--$\pi^{\ast}$
%exciton-dominated
%electronic transitions in graphene
%cause a strong absorption band 
%near 2755$\Angstrom$ 
%(Yang et al.\ 2009, Nelson et al.\ 2010, 
%Trevisanutto et al.\ 2010)
%which is not seen in the ISM.
%
%In this work, we aim at placing 
%a quantitative upper limit on
%the abundance of interstellar graphene
%on the basis of the nondetection 
%of the fingerprint 2755$\Angstrom$ 
%absorption feature in the diffuse ISM.
%
%To achieve this, we first calculate
%in \S\ref{sec:extcurv} 
%the UV absorption of graphene 
%and compare it with the Galactic
%interstellar extinction curve.
%
%Also, if graphene is present in the diffuse ISM,
%single-photon heating by starlight 
%(Draine \& Li 2001)
%will cause it to radiate in the IR
%through its characteristic vibrational transitions
%(e.g., see Garc{\'{\i}}a-Hern{\'a}ndez et al.\ 2011b,
%Mackie et al.\ 2015).
%Therefore, we calculate in \S\ref{sec:irem}
%the IR emission spectrum of graphene 
%heated by starlight, 
%and compare it with measurements of 
%the IR emission of the diffuse ISM 
%by the {\it Infrared Telescope in Space}
%(IRTS; Onaka et al.\ 1996) and by
%the {\it Diffuse Infrared Background 
%Experiment} (DIRBE; Arendt et al.\ 1998) 
%and the {\it Far Infrared Absolute
%Spectrophotometer} (FIRAS; 
%Finkbeiner et al.\ 1999) instruments on 
%the {\it Cosmic Background Explorer} 
%(COBE) satellite.
%The major conclusion is summarized 
%in \S\ref{sec:summary}.
%

\section{Extinction}\label{sec:extcurv}
If we approximate a 2-dimensional graphene monolayer
as a very flat oblate (``pancake''), in the Rayleigh regime
the absorption cross section 
$C_{\rm abs}(\lambda)$ at wavelength $\lambda$ 
per unit volume ($V$) is
\begin{equation}\label{eq:Cabs2V}
C_{\rm abs}(\lambda)/V = \frac{2\pi}{\lambda}
\times {\rm Im}\left\{\frac{\varepsilon -1}
{1 + L \left(\varepsilon-1\right)}\right\},
\end{equation}
where $\varepsilon(\lambda)$ is the complex
dielectric function in the graphene plane,\footnote{%
   For a graphene monolayer, the dielectric function 
  $\varepsilon$ along the c-axis perpendicular
  to the graphene plane is close to 1 in the wavelength
  range of interest here (see Nelson et al.\ 2014). 
  Therefore, the contribution to $C_{\rm abs}$ 
  by the dielectric function along the c-axis 
  is negligible.
  }
$L\approx0$ is the depolarization factor
for the electric field vector $\vec{E}$ 
along the graphene plane, and
${\rm Im\left\{...\right\}}$ 
denotes the imaginary part 
of a complex function
(see eq.\,3.40 of Kr\"ugel 2003).
By relating the graphene volume $V$ to $\NC$,
the number of C atoms 
contained in the graphene sheet
(see eq.\,4 of Chen at al.\ 2017),
we calculate $C_{\rm abs}(\lambda)$  
from $\varepsilon$ as follows:
\begin{equation}\label{eq:Cabs2NC}
C_{\rm abs}(\lambda)/\NC = \frac{24\pi\,m_{\rm H}\,d}
{\sigma\lambda}\,{\rm Im}\left\{\varepsilon-1\right\} ~~,
\end{equation}
where $m_{\rm H}\approx1.66\times10^{-24}\g$ 
is the mass of a hydrogen nuclei,
$d\approx3.4\Angstrom$ is the monolayer
thickness of graphene, and
$\sigma\approx6.5\times10^{-8}\g\cm^{-2}$
is the surface mass density of graphene.

While Chen et al.\ (2017) adopted the SE-measured 
dielectric functions of Nelson et al.\ (2010),
in this work we will use the EELS-measured
dielectric functions of Nelson et al.\ (2014).\footnote{%
   When deducing dielectric functions from 
   the experimental energy loss sepctrum of graphene,
   Nelson et al.\ (2014) treated graphene as isotropic, 
   although it is surely anisotropic. 
   However, as shown by the theoretical 
   energy loss spectrum of Nelson et al.\ (2014)
   calculated from first-principles methods 
   based on density functional theory (DFT), 
   the experimental spectrum is dominated 
   almost entirely by electronic excitations 
   with momentum transfer parallel to 
   the basal plane, i.e., the graphene layer
   (see the lower panel of Figure~3b in
   Nelson et al.\ 2014). The imaginary parts
   of the DFT-calculated dielectric functions 
   along the ``c-axis''  (which is normal to 
   the basal plane) 
   are negligible at $\ltsim$\,10$\eV$
   compared to that along the basal plane.
   }
%
%
% ... the "c-axis" $\hat{{\boldsymbol{c}}}$ 
% is normal to the basal plane (i.e., the graphene layers). 
%The dielectric tensor has two components: 
%${\epsilon }_{\parallel }(\omega )$ 
%describing the response to electric fields 
%${\boldsymbol{E}}\parallel \hat{{\boldsymbol{c}}}$, 
%and ${\epsilon }_{\perp }(\omega )$ for the response 
%when ${\boldsymbol{E}}\perp \hat{{\boldsymbol{c}}}$.
%
%
%
As mentioned in \S\ref{sec:intro},
the graphene film measured with SE was
exposed to air and the majority of its surface 
was covered with adsorbates
and contamination that would affect the measurement 
of its optical properties (see Nelson 2012).
In contrast, the EELS method could allow one 
to localize the measurement to an area that is 
free of contamination or the effects of a substrate. 
In this way the EELS technique could give a {\it purer} 
dielectric response than an optical measurement like SE 
in terms of completely isolating the graphene monolayer.

Also, while the SE method measures the dielectric functions 
of graphene in the energy range of $\simali$0.7--5$\eV$
(which corresponds to a wavelength range of 
$0.153\mum\simlt \lambda \simlt 0.783\mum$
or $1.28\mum^{-1} \simlt \lambda^{-1} \simlt6.54\mum^{-1}$),
the EELS method probes a much {\it wider} energy range 
at $\simali$0--50$\eV$ 
(i.e., $\lambda\simgt0.025\mum$
or $\lambda^{-1} \simlt 40\mum^{-1}$).
The EELS dielectric functions are astrophysically
more relevant since the Galactic interstellar 
extinction curve spans a wavelength range 
much {\it wider} than that probed 
by SE but well {\it within} that of EELS.
As demonstrated in Chen et al.\ (2017), 
to facilitate a direct comparison of
the absorption of graphene 
with the interstellar extinction curve, 
an extrapolation of the SE-measured 
dielectric functions (Nelson et al.\ 2010) 
over a wide wavelength range was made 
by fitting the SE-measured dielectric functions 
with three Lorentz oscillators.
A major advantage of the EELS method 
over SE is that no extrapolation is needed
for the EELS-measured dielectric functions.

%The main advantages of the EELS method over SE are the
%abilibites to probe a much wider energy range 
%($\simali$0--50\,eV) as well as localize the measurement
%to an area that is free of contamination or the effects of 
%a substrate. In this way EELS can give a purer dielectric 
%response than an optical measurement can in terms
%of completely isolating the graphene monolayer.
%
%The previous optical studies of graphene from SE
%measurements made in air, 
%in which graphene is sitting on a substrate,
%have used a 3.35$\Angstrom$ thickness based 
%on the interlayer spacing of bulk graphite. 

\begin{figure*}
\centering
\includegraphics[width=8cm, angle=0]{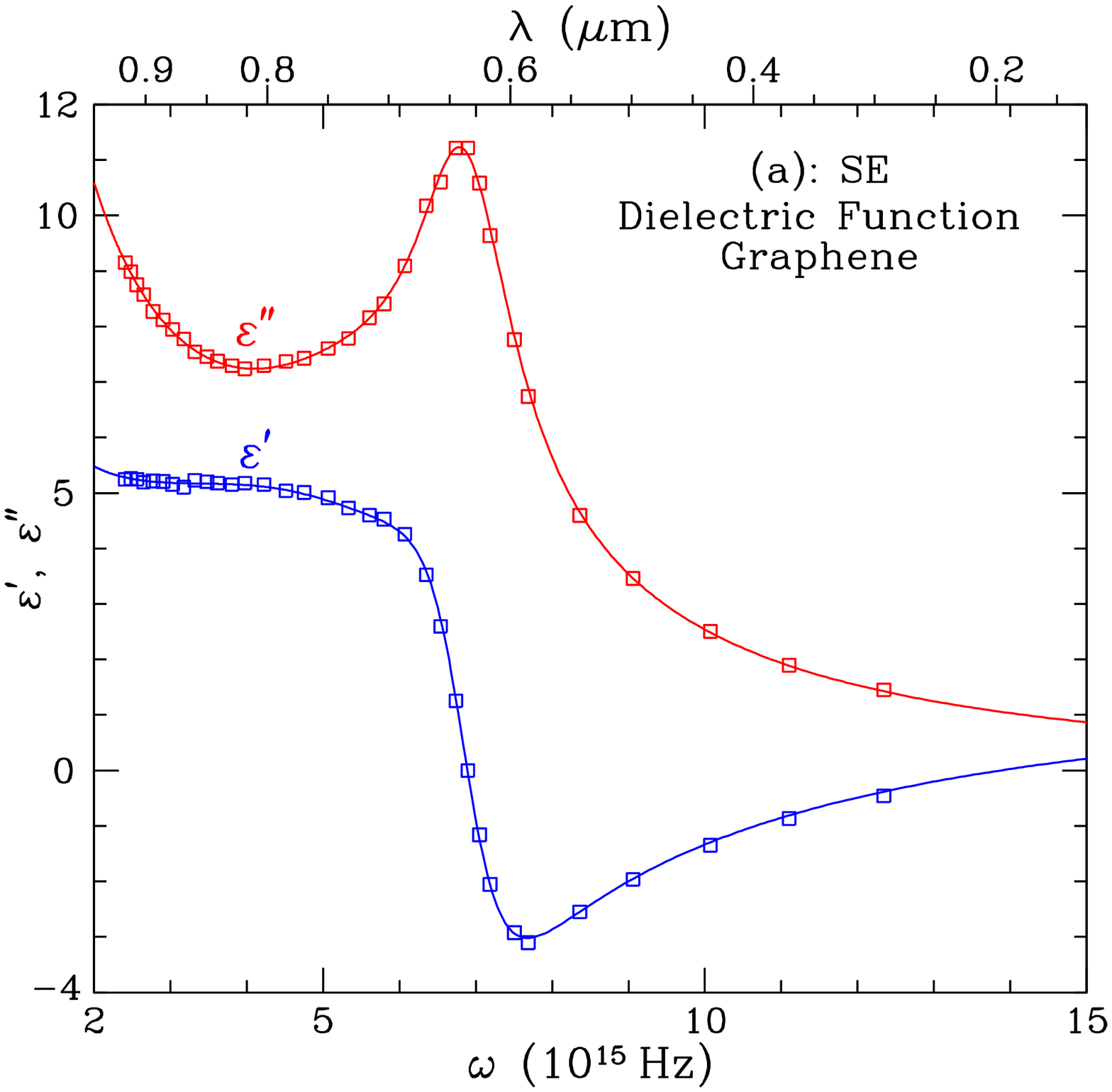}
\includegraphics[width=8cm, angle=0]{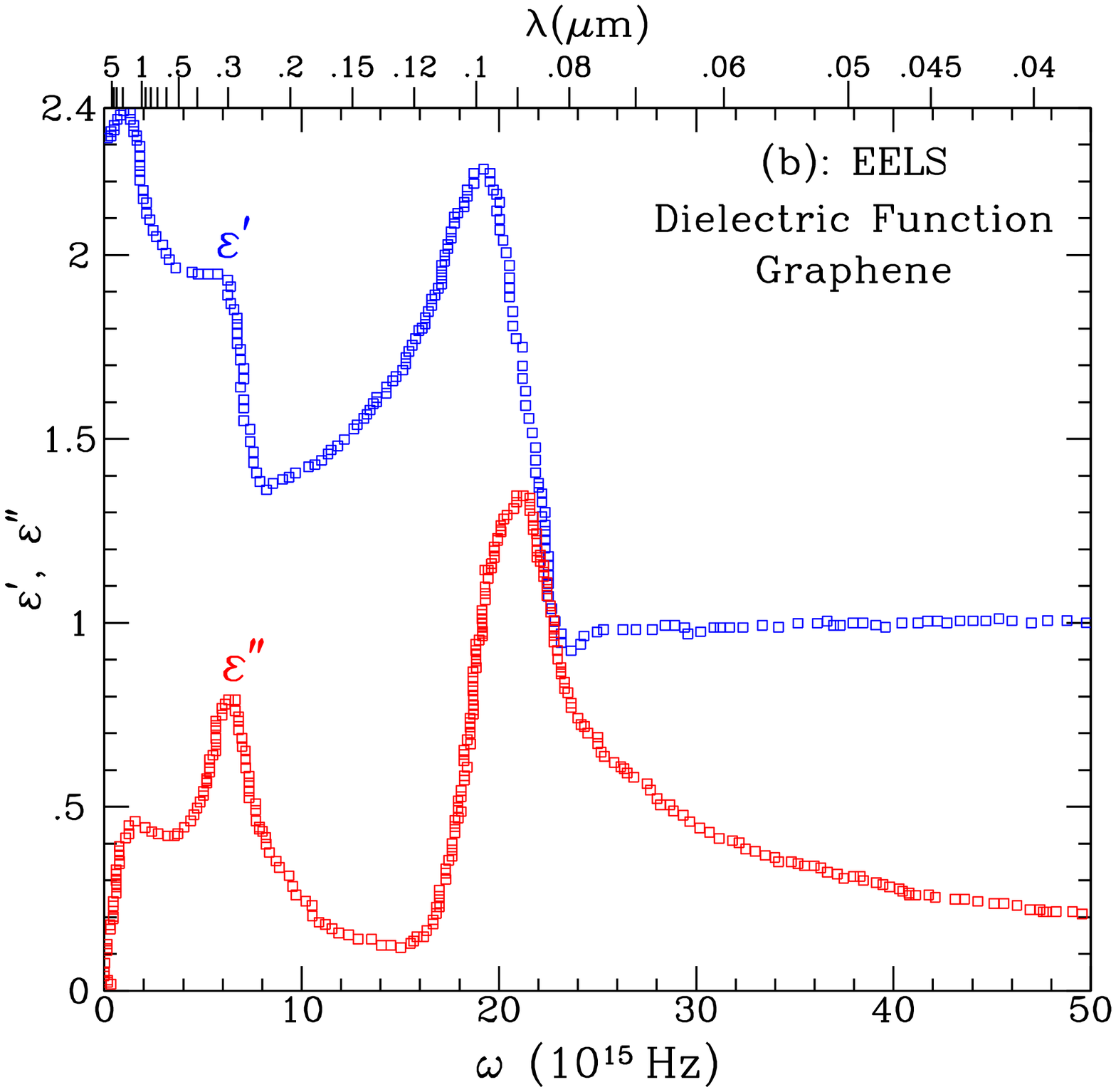}
\caption{\label{fig:dielfunc}
         Real ($\epsre$) and imaginary ($\epsim$)
         parts of the dielectric functions of graphene
         as measured by the SE method (a) and
         by the EELS method (b). 
         Open squares: experimental data of 
         Nelson et al.\ (2010) for SE
         and of Nelson et al.\ (2014) for EELS.
         Solid lines: model fits of Chen et al.\ (2017)
         to the SE-measured data with three Lorentz oscillators. 
         }
\end{figure*}

A comparison of the SE vs. EELS dielectric functions 
is shown in Figure~\ref{fig:dielfunc}. 
Most notably, both the real ($\epsre$) and imaginary
($\epsim$) parts of the EELS dielectric functions are
substantially lower than that of SE.\footnote{%
   For $\epsre$, it is actually $|\epsre|$,
   the absolute value of SE that always exceeds 
   that of EELS since $\epsre$ beccomes negative
   at some wavelengths.
   } 
%Also, the $\pi$--$\pi^\ast$ electronic interband transition
%seen at $\simali$4.5$\eV$ 
%(i.e, $\lambda\approx2755\Angstrom$,
%$\lambda^{-1}\approx3.63\mum^{-1}$)
%in the SE measurement shifts to a higher energy
%at $\simali$4.9$\eV$ 
%(i.e, $\lambda\approx2534\Angstrom$,
%$\lambda^{-1}\approx3.95\mum^{-1}$).
The difference between the SE and EELS 
dielectric functions of graphene clearly 
demonstrates that the presence of 
surface contamination in the SE method
(due to measurement in air) could considerably
affect the measurement of its optical properties.
We also note that, as shown in Nelson et al.\ (2014),  
the experimentally measured EELS spectrum of 
graphene closely agrees with that calculated 
from the two-dimensional hydrodynamic model
of Jovanovic et al.\ (2011).

\begin{figure*}
\centering
\includegraphics[width=8cm]{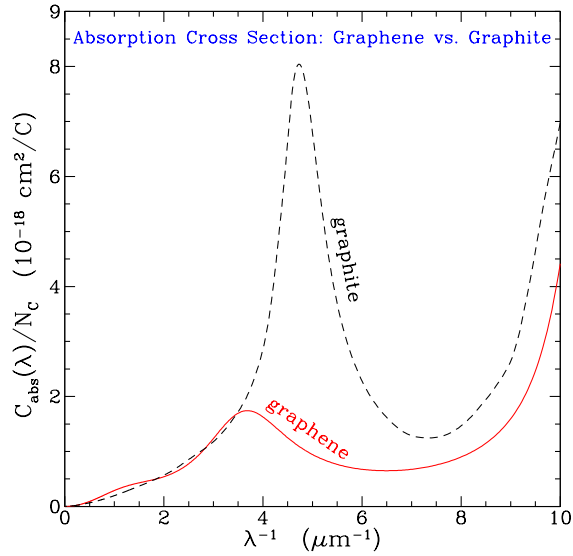}
\caption{\label{fig:Cabs}
         UV/optical absorption cross sections (per C atom)
         of graphene (solid line) 
         and nano-sized graphite (dashed line). 
         }
\end{figure*}

In Figure~\ref{fig:Cabs} we show the UV/optical 
absorption cross section (per C atom) of graphene
which exhibits a gradual increase from the near-IR
through the visible to the near-UV and
a prominent absorption band peaking 
at $\simali$4.5$\eV$ 
(i.e, $\lambda\approx2755\Angstrom$,
$\lambda^{-1}\approx3.63\mum^{-1}$),
followed by a relatively flat plateau
at $\simali$6--8$\mum^{-1}$ 
and a steep rise at $\lambda\gtsim8\mum^{-1}$.
The $\pi$--$\pi^\ast$ electronic interband transition
peak at $\simali$4.5$\eV$ 
has been predicted by quantum-chemical calculations 
(see Yang et al.\ 2009, Trevisanutto et al.\ 2010)
and seen in the SE spectrum of the CVD graphene
(Nelson et al.\ 2010). While it was previously ascribed 
to plasmons, Nelson et al.\ (2014) attributed it to
single-particle interband excitations.
Also shown in Figure~\ref{fig:Cabs} is 
the extinction cross section of a graphite
nano particle of $\NC=40$ 
calculated from Mie theory 
(Bohren \& Huffman 1983)
using the dielectric functions of 
``astronomical graphite'' 
of Draine \& Lee (1984).\footnote{%
   Nano-sized graphite grains are in 
   the Rayleigh regime even in the far-UV
   wavelength range . Hence the extinction
   is dominated by absorption while the scattering
   is negligible (see Bohren \& Huffman 1983).
   } 
At $\lambda^{-1}\simlt3.5\mum^{-1}$,
the absorption cross section (per unit mass)
of graphene agrees remarkably well with
that of graphite. The major difference lies
at $\lambda^{-1}\simgt3.5\mum^{-1}$:
while graphene peaks at 
$\simali$2755$\Angstrom$,
graphite exhibits a prominent absorption 
peak at $\simali$2175$\Angstrom$
(i.e., $\simali$4.60$\mum^{-1}$).\footnote{%
     It is interesting to note that the absorption
     cross sections of graphite calculated from 
     the dielectric functions newly compiled 
     by Draine (2016) for the electric field vector 
     parallel to the c-axis also exhibit
     an absorption band at $\simali$2755$\Angstrom$
    (see Figure~10 of Draine 2016).
    }

\begin{figure*}
\centering
\includegraphics[width=.5\textwidth]{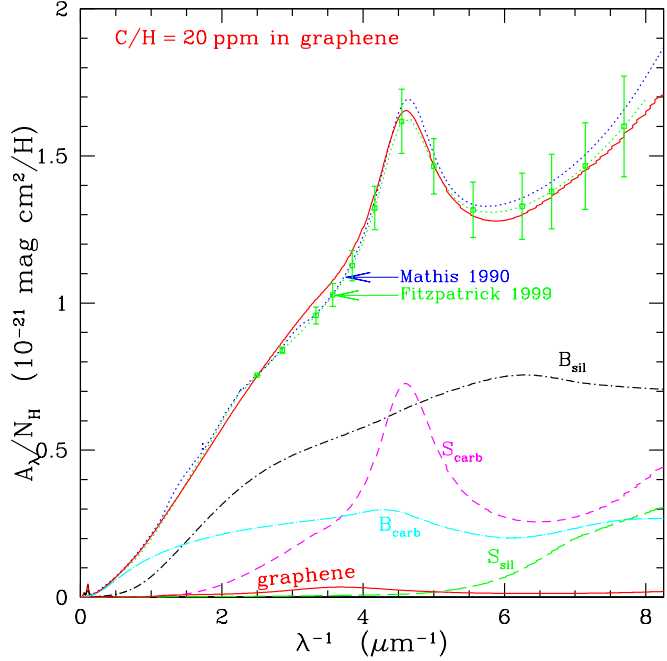}
\vspace{20mm}
\caption{\label{fig:extcurv}
        Comparison of the average Galactic interstellar extinction 
        curve (open circles: Mathis 1990; dotted line: Fitzpatrick 
        1999) with the model extinction curve (solid red line)
        obtained by adding the contribution from graphene
        with $\CTOHgraphene=20\ppm$ (thin red line) 
        to the best-fit model of Weingartner \& Draine (2001).
        Also plotted are the contributions 
        (see Li \& Draine 2001b) from 
        ``${\rm B_{sil}}$'' ($a \ge 250$\AA\ silicate);
        ``${\rm S_{sil}}$'' ($a < 250$\AA\ silicate); 
        ``${\rm B_{carb}}$'' ($a\ge 250$\AA\ carbonaceous); 
        ``${\rm S_{carb}}$'' ($a < 250$\AA\ carbonaceous, 
        including PAHs). 
        The vertical bars superposed on the extinction curve
        of Fitzpatrick (1999) represent the observational uncertainties.       
        }
\end{figure*}

As shown in Figure~\ref{fig:extcurv},
the Galactic extinction curve exhibits
a strong absorption band at 2175$\Angstrom$
but lacks any spectral features around 
2755$\Angstrom$. As demonstrated 
in Chen et al.\ (2017), the nondetection of 
the 2755$\Angstrom$ absorption feature 
in the extinction curve allows us to 
place an upper limit on the abundance 
of graphene in the ISM.
Following Chen et al.\ (2017),
we achieve this by adding graphene
to the silicate-graphite-PAH model
of Weingartner \& Draine (2001) 
and Li \& Draine (2001b)
which reproduces both the observed interstellar
extinction curve and the observed IR emission.
Let $\CTOHgraphene$ be the amount of 
C (relative to H) tied up in graphene.
The extinction results from graphene 
of a quantity of $\CTOHgraphene$ is
\begin{equation}
\left(\frac{A_\lambda}{\NH}\right)_{\rm graph}
= 1.086\,\left(\frac{C_{\rm abs}}{\NC}\right)_{\rm graph}
\CTOHgraphene ~~.
\end{equation}
%
%
%%
%In Figure~\ref{fig:extcurv}
%we show the extinction obtained 
%by {\it adding} graphene to this model.
%Since graphene is in the Rayleigh limit, 
%the added extinction depends only on 
%$\CTOHgraphene$, the amount of C 
%(relative to H) tied up in graphene, 
%and is independent of the exact graphene size:
%%
%\begin{equation}
%\left(\frac{A_\lambda}{\NH}\right)_{\rm graph}
%= 1.086\,\left(\frac{C_{\rm abs}}{\NC}\right)_{\rm graph}
%\CTOHgraphene ~~.
%\end{equation}
%%
%
As shown in Figure~\ref{fig:extcurv}, 
the maximum amount of graphene 
allowable in the ISM is derived by requiring
the graphene-added model extinction
not to exceed the observational uncertainties
of the Galatcic interstellar extinction curve 
(Mathis 1990, Fitzpatrick 1999).
In this way, we estimate the upper bound to 
be $\CTOHgraphene\approx20\ppm$,
%i.e., in the diffuse ISM there could 
%be as much as $\simali$20$\ppm$ of C/H 
%in graphene for the characteristic 
%2755$\Angstrom$ absorption feature 
%of graphene to remain unnoticeable. 
%This upper limit is derived by requiring
%the graphene-added model extinction
%not to exceed the observational uncertainties
%of the interstellar extinction curve of Fitzpatrick (1999).
%
%The contribution of graphene 
%to the $\lambda^{-1}\sim3.6\mum^{-1}$ region 
%would be appreciable for $\CTOHgraphene>20\ppm$.
corresponding to $\simali$7\% of 
the total interstellar C,
if the interstellar C abundance is solar
(i.e., C/H\,$\approx$\,269$\pm$31$\ppm$,
Asplund et al.\ 2009).
%, an upper limit of
% $\CTOHgraphene\approx20\ppm$
%implies that at as much as $\simali$7\% of 
%the interstellar C atoms 
%could be tied up in graphene.
We note that the upper limit of $\simali$20$\ppm$
derived here is higher by a factor of $\simali$3
than that of Chen et al.\ (2017).
This is expected since the dielectric functions
of graphene adopted here were measured with EELS
for clean graphene, while Chen et al.\ (2017) adopted
the dielectric functions of contaminated graphene
measured with SE in air. 
As shown in Figure~\ref{fig:dielfunc},
the former are considerably smaller than the latter. 
We note that, as shown in eq.\,\ref{eq:Cabs2NC},
the UV/optical absorption (per unit mass)
of graphene linearly increases with $\epsim$, 
the imaginary parts of the dielectric functions of graphene.

\section{Emission}\label{sec:irem}
Graphene, if indeed present in the ISM,  
would absorb UV/optical stellar photons
and re-radiate the absorbed photon energy 
in the IR. Therefore, graphene could also
reveal its presence in the ISM through its
characteristic C--C vibrational bands. 
Indeed, C$_{24}$, a small graphene sheet,
has been possibly detected in several 
Galactic and extragalactic PNe
(Garc{\'{\i}}a-Hern{\'a}ndez et al.\ 2011a, 2012)
and even in the ISM (Bern\'e et al.\ 2013)
through its characteristic bands
at $\simali$6.6, 9.8 and 20$\mum$.
Chen et al.\ (2017) modeled the vibrational 
excitation and radiative relaxation of C$_{24}$
in two interstellar regions 
--- the high Galactic-latitude cirrus 
and the diffuse ISM toward
$l = 44^{\rm o}20^\prime$, 
$b=-0^{\rm o}20^\prime$.
The former is illuminated by
the solar neighbourhood interstellar
radiation field (ISRF) estimated by 
Mathis et al.\ (1983; hereafter MMP83).
The starlight in the latter region 
can also be approximated by 
the solar neighbourhood ISRF 
but with an intensity twice as 
that of MMP83 (see Li \& Draine 2001b).

\begin{figure*}
\centering
\includegraphics[width=.4\textwidth]{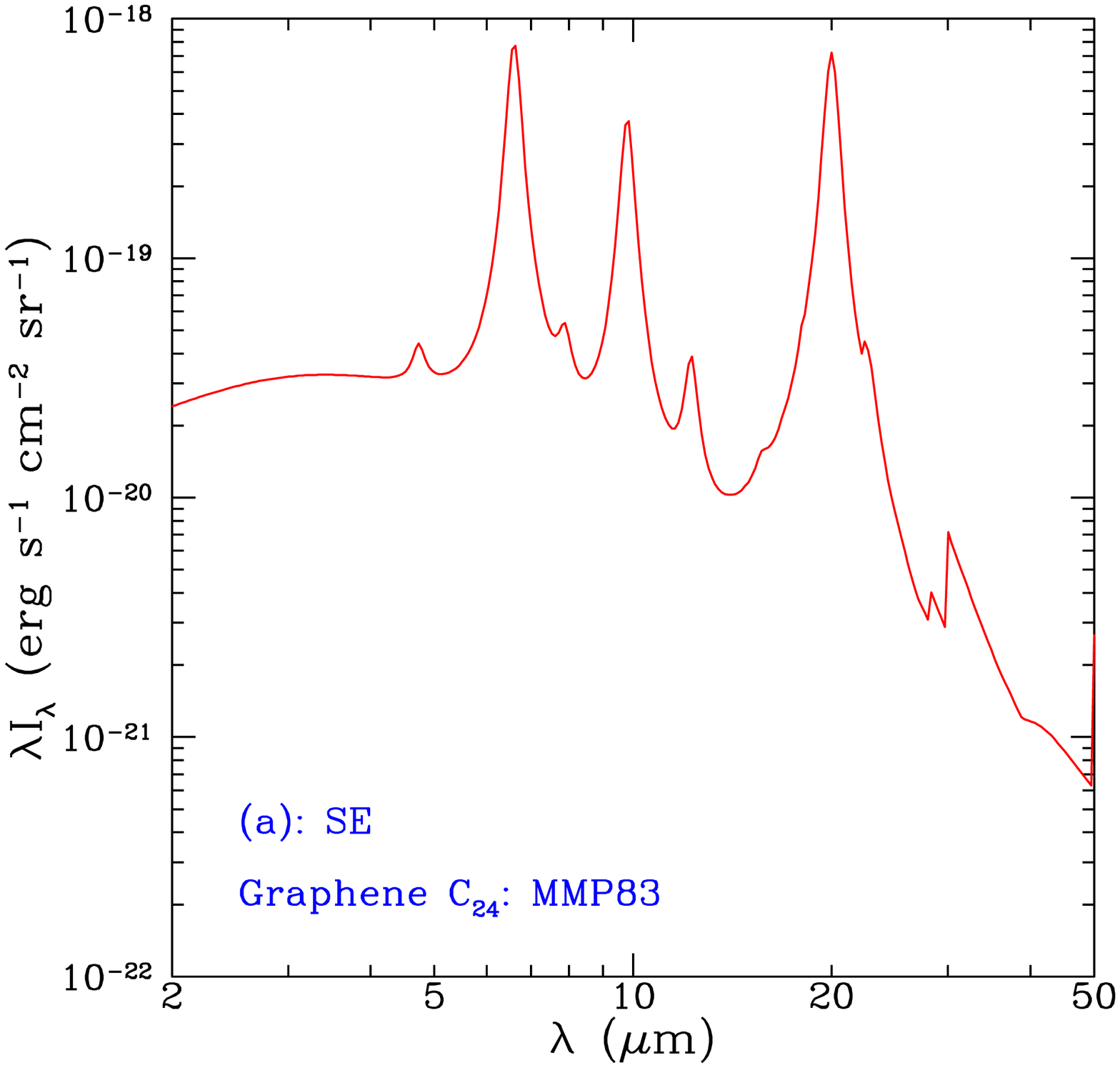}
\includegraphics[width=.4\textwidth]{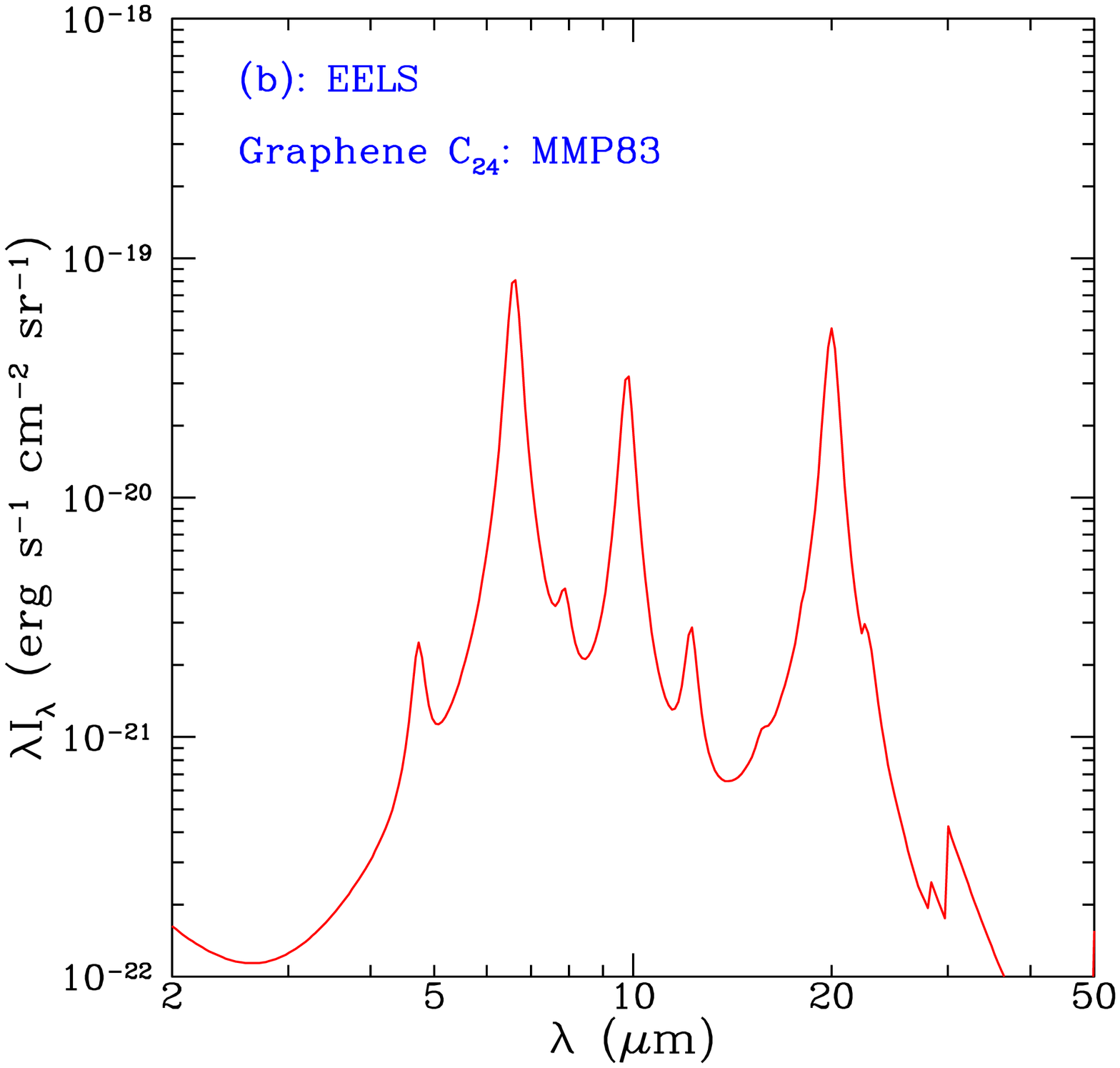}
\caption{\label{fig:irem}
         IR emission spectrum 
         of graphene of $\NC=24$
         illuminated by the MMP83 ISRF.
         Left panel (a): the dielectric functions
         of Nelson et al.\ (2010)
         measured with the SE method 
         were adopted.
         Right panel (b): the dielectric functions
         of Nelson et al.\ (2014)
         measured with the EELS method 
         were adopted.
         %The sawtooth features 
         %at $\lambda>30\mum$ are
         %due to our treatment of 
         %transitions from the lower 
         %excited energy bins to 
         %the ground state and first 
         %few excited states
         %(see Draine \& Li 2001).
         }
\end{figure*}

Using the UV/optical absorption cross section
derived from the SE dielectric functions 
of Nelson et al.\ (2010), 
Chen et al.\ (2017) calculated the IR
emission spectrum of C$_{24}$ excited
by the MMP83 ISRF. 
As illustrated in Figure~\ref{fig:irem}a,
the 6.6, 9.8 and 20$\mum$ features
are prominent in the model emission
spectrum of C$_{24}$.
However, these emission features
are neither seen in the Galactic cirrus 
nor in the diffuse ISM toward
$l = 44^{\rm o}20^\prime$, 
$b=-0^{\rm o}20^\prime$.  
Chen et al.\ (2017) found that the nondetection
of the fingerprint bands of C$_{24}$
at $\simali$6.6, 9.8 and 20$\mum$
is consistent with an upper limit of
$\simali$5$\ppm$ of C$_{24}$ in these regions.

As discussed in \S\ref{sec:extcurv},
the true UV/optical absorption of graphene
would be considerably weaker than 
that derived in Chen et al.\ (2017).
By implication, graphene would emit
much less (on a per unit mass basis)
compared to that calculated by Chen et al.\ (2017).
Therefore, we expect to derive 
a higher upper limit on the abundance of C$_{24}$ 
from the nondetection of the C$_{24}$ bands
at $\simali$6.6, 9.8 and 20$\mum$
in the Galactic cirrus 
and in the diffuse ISM toward
$l = 44^{\rm o}20^\prime$, 
$b=-0^{\rm o}20^\prime$.  

Following Chen et al.\ (2017),
we employ the ``exact-statistical'' method
developed by Draine \& Li (2001) to model
the stochastic heating of C$_{24}$.\footnote{%
   For a planar graphene of several hundred 
   C atoms or smaller, its energy content is 
   often smaller than the energy of a single 
   UV/optical stellar photon. 
   Therefore, graphene will not attain
    an equilibrium temperature but undergo 
    stochastic heating in the ISM 
    (Greenberg 1968).
    }
We use the vibrational modes and intensities
of the C$_{24}$ graphene obtained by Martin et al.\ (1996) 
and Kuzmin \& Duley (2011) from DFT-based 
quantum-chemical computations. 
Due to lack of data on the UV/optical absorption 
of C$_{24}$, we adopt that of graphene derived from 
the EELS-measured dielectric functions
(see Figure~\ref{fig:Cabs}),
although the actual absorption spectrum
of C$_{24}$ may not be as smooth 
as the EELS-derived data and may have
strong, sharp individual UV absorption features.
For illustration, we show in Figure~\ref{fig:irem}b
the model emission spectrum of C$_{24}$ 
excited by the MMP83 ISRF. As expected,
the model emission intensity of  C$_{24}$ 
calculated here is substantially lower
than that of Chen et al.\ (2017).

\begin{figure*}
\centering
\includegraphics[width=.36\textwidth, angle=270]{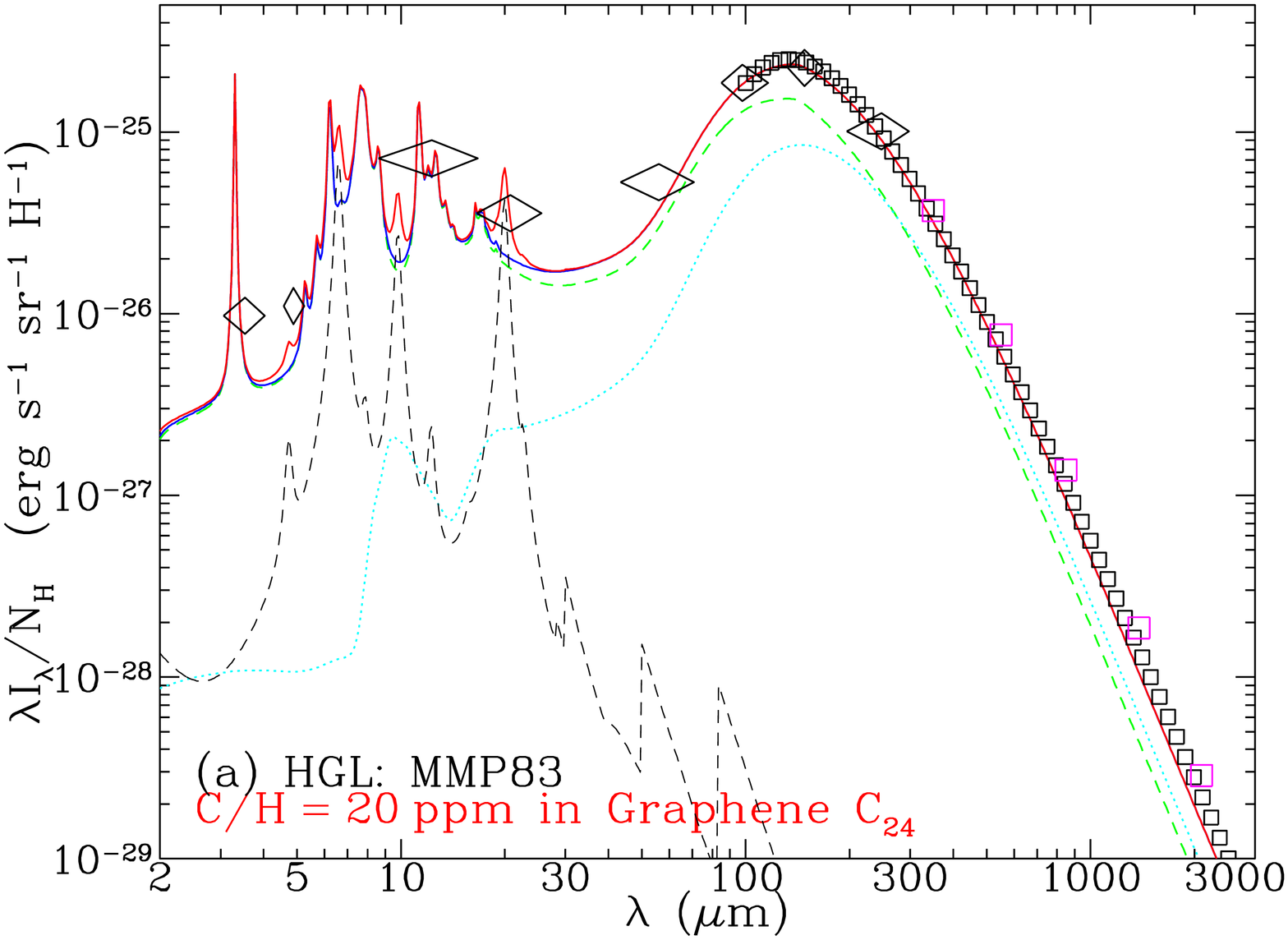}
\includegraphics[width=.36\textwidth, angle=270]{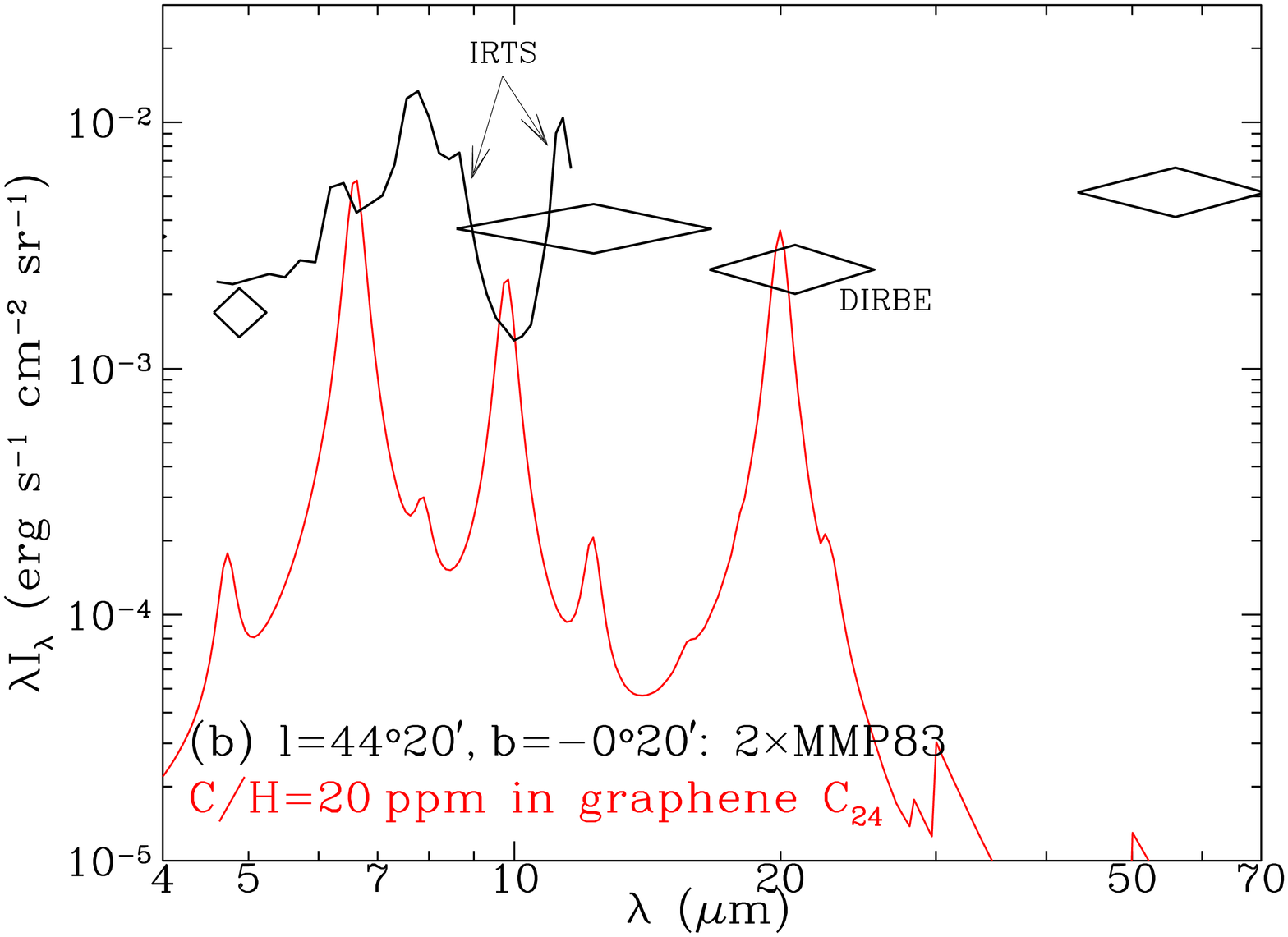}
\caption{\label{fig:dism}
        Left panel (a):
        Comparison of the observed IR emission 
        from the HGL cirrus
        with the model emission spectrum 
        obtained by adding the contribution 
        from graphene C$_{24}$
        of $\CTOHgraphene=20\ppm$ (black dashed line) 
        to the best-fit model of Li \& Draine (2001b)
        which consists of contributions from
        amorphous silicate (cyan dotted line) and
        carbonaceous grains (i.e., graphite and PAHs;
        green dashed line). The sum of amorphous silicate 
        and carbonaceous grains is shown as blue solid line.
        Observational data are from 
        {\it DIRBE} (black diamonds; Arendt et al.\ 1998), 
        {\it FIRAS} (black squares; Finkbeiner et al.\ 1999), 
        and {\it Planck} (magenta squares; 
        Planck Collaboration XVII 2014).
        Right panel (b):
        Contribution to the IR emission toward
        ($44^{\rm o}\le l \le 44^{\rm o}40^\prime$, 
        $-0^{\rm o}40^\prime\le b \le 0^{\rm o}$) by
        graphene C$_{24}$
        of $\CTOHgraphene=20\ppm$ (solid red line). 
	Diamonds: {\it DIRBE} photometry.
	Black solid line:
        5--12$\mum$ spectrum observed by IRTS
        (Onaka et al.\ 1996). 
        }
\end{figure*}

Similar to Chen et al.\ (2017),
we also derive upper limits on 
the abundance of the graphene C$_{24}$
in the Galactic cirrus and 
in the diffuse ISM toward
$l = 44^{\rm o}20^\prime$, 
$b=-0^{\rm o}20^\prime$
based on comparison of the observed 
IR emission with the calculated emission 
spectrum of C$_{24}$.
For the Galactic cirrus, 
the average emission per H 
has been measured by {\it COBE}/DIRBE 
(Arendt et al.\ 1998), {\it COBE}/FIRAS 
(Finkbeiner et al.\ 1999), and {\it Planck} 
(Planck Collaboration XVII 2014). 
The diffuse ISM toward
$l = 44^{\rm o}20^\prime$, 
$b=-0^{\rm o}20^\prime$
has been observed
by {\it COBE}/DIRBE
(Hauser et al.\ 1998).
The {\it Mid-Infrared Spectrograph} (MIRS) 
aboard the {\it Infrared Telescope in Space}
(IRTS) has also obtained the 4.7--11.7$\mum$ 
spectrum for the diffuse ISM toward
$l = 44^{\rm o}20^\prime$, 
$b=-0^{\rm o}20^\prime$ (Onaka et al.\ 1996).
As shown in Figure~\ref{fig:dism}a,
even locking up 20$\ppm$ of C/H
--- the upper limit of graphene
derived from the interstellar extinction
--- all in the specific graphene species C$_{24}$,
the 6.6, 9.8 and 20$\mum$ emission features
of C$_{24}$ would still be hidden by the PAH features 
at 6.2, 7.7, 8.6 and 11.3$\mum$ 
and would remain undetected by {\it Spitzer} or 
by the {\it Short Wavelength Spectrometer} (SWS)
aboard the {\it Infrared Space Observatory} (ISO). 
This is also true for
the diffuse ISM toward
$l = 44^{\rm o}20^\prime$, 
$b=-0^{\rm o}20^\prime$. 
As illustrated in Figure~\ref{fig:dism}b,
as much as $\simali$20$\ppm$ of C/H
could also be tied up in the C$_{24}$ graphene 
while the characteristic 6.6, 9.8 and 20$\mum$ 
emission features of C$_{24}$ are still not strong 
enough to be detected by {\it IRTS}.
Therefore, for both the Galactic cirrus
and the line of sight toward 
$l = 44^{\rm o}20^\prime$, $b=-0^{\rm o}20^\prime$,
a upper limit of C/H\,$\simlt$\,20$\ppm$
is imposed by the {\it COBE}/DIRBE photometric data
and the {\it IRTS} spectrum. 
Nevertheless, the actual abundance of C$_{24}$ 
could be much lower than 20$\ppm$ since,
if graphene is indeed present in the ISM, it could
span a wide range of sizes and charging states.

\section{Summary}\label{sec:summary}
We have explored the possible presence of
graphene in the diffuse ISM based on the dielectric
functions of graphene recently measured with
the EELS method. 
Our principal results are as follows:
\begin{enumerate}
\item Using the EELS dielectric functions of
          graphene, we have calculated its UV/optical 
          absorption and compared it with the observed
          interstellar extinction curve. 
\item Based on the absence of
          the $\simali$2755$\Angstrom$ 
          $\pi$--$\pi^{\ast}$ electronic
          absorption band of graphene
          in the interstellar extinction curve, 
          we have placed an upper limit of 
          $\simali$20$\ppm$ of C/H
          (i.e., $\simali$7\% of the total interstellar C)
          on the interstellar graphene abundance.
\item The upper limit of $\simali$20$\ppm$ of C/H
          on graphene derived here with the EELS dielectric
          functions exceeds the previous SE-based estimate
          by a factor of $\simali$3.
          This is because the SE method measures
          the dielectric functions of graphene exposed to air
          and the majority of its surface would be covered 
          with adsorbates and contamination,
          while the EELS method localizes the measurement
          to an area that is free of contamination.
          As a result, the EELS dielectric functions are
          substantially smaller than the SE dielectric functions 
          and thus the UV/optical absorption of graphene
          (on a per unit mass basis)
          calculated here using the EELS dielectric functions 
          is considerably smaller than that of SE.
\item Inspired by the possible detection of
         the 6.6, 9.8 and 20$\mum$ emission features
         of the C$_{24}$ graphene in several Galactic 
         and extragalactic PNe, we have also explored 
         the possible presence of this specific graphene
         species in two interstellar regions ---
         the high Galactic latitude cirrus and 
         the diffuse ISM toward 
         $l = 44^{\rm o}20^\prime$, 
         $b=-0^{\rm o}20^\prime$ ---
         by modeling the vibrational excitation 
         and radiative emission processes of C$_{24}$.
         Using the EELS dielectric functions of graphene
         to calculate the UV/optical absorption of C$_{24}$,         
         we have found that as much as $\simali$20$\ppm$
         of C/H could be tied up in C$_{24}$
         while the 6.6, 9.8 and 20$\mum$ emission 
         features of C$_{24}$ would remain undetected 
         in these two interstellar regions
         by {\it IRTS}, {\it ISO}/SWS, or {\it Spitzer}.
\item We note that the true interstellar abundance 
          of C$_{24}$ could be much lower than 20$\ppm$ 
          since, if graphene is indeed present in the ISM, 
          it could span a wide range of sizes and charging states.
          We call for further quantum-chemical computations 
         and experimental measurements
         of the IR vibrational spectra of graphene
         of a wide range of sizes and their cationic 
         as well as  anionic counterparts.
\end{enumerate}

\section*{Acknowledgements}
We thank X.H~Chen, A.~Diebold, 
B.T.~Draine, W.W.~Duley,
%D.A.~Garc{\'{\i}}a-Hern{\'a}ndez,
Th.~Hening, C.~J\"ager, S.~Kuzmin, 
S.K.~Madhulika,  L.~Yang
and the anonymous referee
for very helpful discussions and suggestions.
This work is supported by NSFC through
Projects 11533002 and 11873041.
AL is supported in part by NSF AST-1816411.

%\clearpage

%%% Figure 1 %%%

%\clearpage

%%% Figure 2 %%%

%\clearpage
%%% Figure 3 %%%

%\clearpage
%%% Figure 4 %%%
%

%\clearpage
%%% Figure 5 %%%

\bsp
\label{lastpage}
\end{document}